\documentclass{sig-alternate}
\usepackage{algorithm}
\usepackage{algorithmic}

\usepackage{tikz}
\usetikzlibrary{calc,trees,positioning,arrows,chains,shapes.geometric,
	decorations.pathreplacing,decorations.pathmorphing,shapes,
	matrix,shapes.symbols,backgrounds}

\usepackage{amsmath, amssymb}
\usepackage{hyperref}
\usepackage{verbatim}
\usepackage{epstopdf}
\usepackage{diagbox}

\newtheorem{theorem}{Theorem}
\newcommand{\SWITCH}[1]{\STATE \textbf{switch} (#1) \begin{ALC@g}}
\newcommand{\ENDSWITCH}{\STATE \textbf{end switch} \end{ALC@g}}
\newcommand{\CASE}[1]{\STATE \textbf{case} #1\textbf{:} \begin{ALC@g}}
\newcommand{\ENDCASE}{\end{ALC@g}}

\newcommand{\DEFAULT}{\STATE \textbf{default:} \begin{ALC@g}}
\newcommand{\ENDDEFAULT}{\end{ALC@g}}
\newcommand{\DEFAULTLINE}[1]{\STATE \textbf{default:} }

\begin{document}
%

\title{
	From Micro to Macro: Uncovering and Predicting Information Cascading Process with Behavioral Dynamics
}
%
%
%
%
%

\numberofauthors{1} 
%
\author{Linyun Yu$^1$, Peng Cui$^1$, Fei Wang$^2$, Chaoming Song$^3$, Shiqiang Yang$^1$\\
	\affaddr{$^1$Department of Computer Science and Technology, Tsinghua University} \\
	\affaddr{$^2$Department of Computer Science and Engineering, University of Connecticut}\\
	\affaddr{$^3$Department of Physics, University of Miami}\\
	\email{yuly12@mails.tsinghua.edu.cn,cuip@tsinghua.edu.cn,fei\_wang@uconn.edu}\\
    \email{chaoming.song@gmail.com, yangshq@tsinghua.edu.cn}
}

\maketitle
\begin{abstract}
Cascades are ubiquitous in various network environments. How to predict these cascades is highly nontrivial in several vital applications, such as viral marketing, epidemic prevention and traffic management. Most previous works mainly focus on predicting the final cascade sizes. As cascades are typical dynamic processes, it is always interesting and important to predict the cascade size at any time, or predict the time when a cascade will reach a certain size (e.g. an threshold for outbreak). In this paper, we unify all these tasks into a fundamental problem: \emph{cascading process prediction}. That is, given the early stage of a cascade, how to predict its cumulative cascade size of any later time? For such a challenging problem, how to understand the micro mechanism that drives and generates the macro phenomenons (i.e. cascading proceese) is essential. Here we introduce behavioral dynamics as the micro mechanism to describe the dynamic process of a node's neighbors get infected by a cascade after this node get infected (i.e. one-hop subcascades). Through data-driven analysis, we find out the common principles and patterns lying in behavioral dynamics and propose a novel Networked Weibull Regression model for behavioral dynamics modeling. After that we propose a novel method for predicting cascading processes by effectively aggregating behavioral dynamics, and propose a scalable solution to approximate the cascading process with a theoretical guarantee. We extensively evaluate the proposed method on a large scale social network dataset. The results demonstrate that the proposed method can significantly outperform other state-of-the-art baselines in multiple tasks including cascade size prediction, outbreak time prediction and cascading process prediction.
\end{abstract}




\section{INTRODUCTION}

In a network environment, if decentralized nodes act on the basis of how their neighbors act at earlier time, these local actions often lead to interesting \emph{macro} dynamics - cascades. In online social networks, the information a user can get and engage in is highly dependent on what his/her friends share, and thus information cascades naturally occur and become the major mechanism for information communication. There has been a growing body of research on these information cascades because of their big potential in various vital applications such as viral marketing, epidemic prevention, and traffic management. Most of them focus on characterizing these information cascades and discovering their patterns in structures, contents and temporal dynamics.

\begin{figure*}
    \centering
    \includegraphics[width=7in]{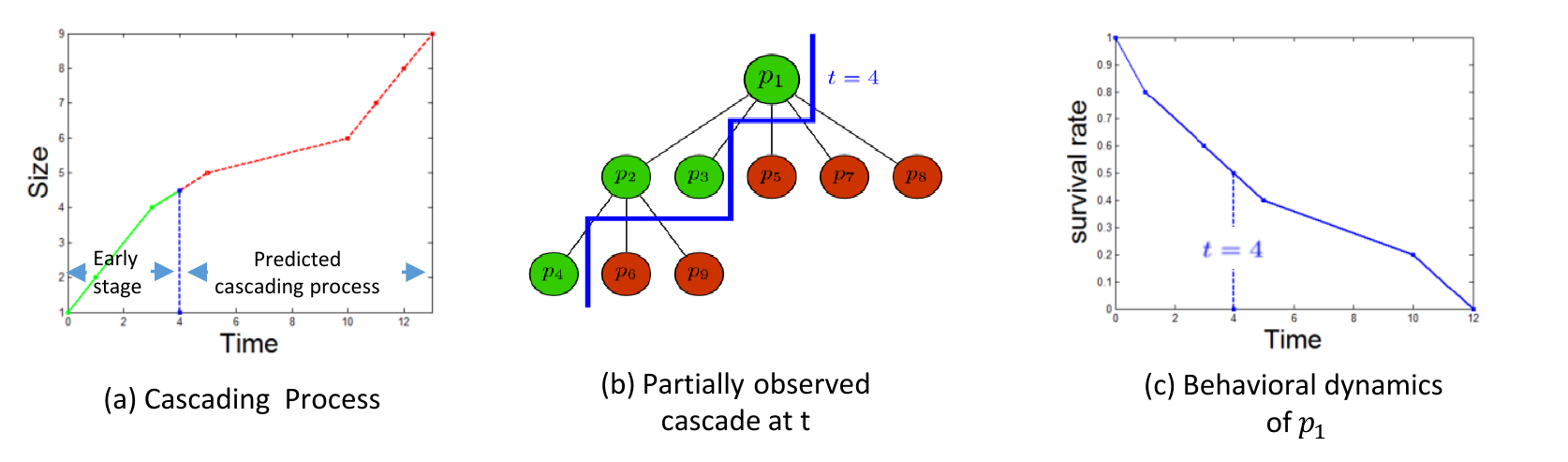}

    \hspace{-0.0002in}
    \vspace{-0.35in}
    \caption{Illustration of cascading process prediction}
    \label{fig:cas_pro}
    \vspace{-0.2in}
\end{figure*}

Recently, predictive modeling on information cascades has aroused considerable research interests. Earlier works on predicting the final size of information cascades based on content, behavioral and structural features \cite{cheng2014can,cui2013cascading}. As only large cascades are of interest in most real applications, Cui et al.\cite{cui2013cascading} propose a data driven approach to predicting whether the final size will surpass a threshold for outbreak. More recently, Cheng et al.\cite{cheng2014can} go beyond the final size to continuously predict whether the cascade will double the current size in future. They also raise an interesting question that whether cascades can be predicted, and their experimental results demonstrate that cascade size are highly predictable. However, the previous works were all about cascade size, which did not include the whole of information cascades. Information cascade is a typical dynamic process, and temporal scale is critical for understanding the cascading mechanism. Also, it is highly nontrivial to predict when a cascade breaks out, and, more ambitiously, to predict the evolving process of a cascade (i.e. cascading process, as shown in \ref{fig:cas_pro} (a)). In this paper, we move one step forward to ask: Is the cascading process predictable? That is, given the early stage of an information cascade, can we predict its cumulative cascade size of any later time?

It is apparent that the targeted problem is far more challenging than those in previous works. The commonly used cascade-level macro features for size prediction, such as the content, increasing speed and structures in the early stage are not distinctive and predictive enough for the cascade sizes at any later time. A fundamental way to address this problem is to look into the \emph{micro} mechanism of cascading processes. Intuitively, an information cascading process can be decomposed into multiple local (one-hop) subcascades. When a node involves in a cascade, one or more of its offspring nodes will also involve in the cascade with a temporal scaling. If the dynamic process of these subcasades can be accurately modeled, then the cascade process can be straightforwardly predicted by an additive function of these local subcascades.

Here we exploit \emph{behavioral dynamics} as the micro mechanism to represent the above mentioned dynamic process of local subcascades. Given a node involving in a cascade at $t_0$, its behavioral dynamic aims at capturing the changing process of the cumulative number of its offspring nodes that involve in the cascade with time evolving. By definition, this is a non-decreasing counting process and can be well represented by survival model \cite{rodriguez2005parametric}. A paucity of recent research works have exploited the survival theory to model how the occurrence of event at a node affects the time for its occurrence at other nodes (i.e. diffusion rate), and their results demonstrate the superiority of continuous-time survival model to uncover temporal processes. However, their targeted problem is to uncover the hidden diffusion networks, and thus suppose the parameters of the survival function on each edge to be fixed. This will cause the unexpected result that all the cascades with the same root node (or early involved nodes) will be anticipated to have the same cascading processes, which makes these models inapplicable in our problem.

In this paper, we propose a novel method for cascading process prediction, as shown in Figure \ref{fig:cas_pro}. Given the early stage of a cascading process before $t$ in Figure \ref{fig:cas_pro} (a), we illustrate the partially observed cascade as shown in Figure \ref{fig:cas_pro} (b), where nodes in green (red) represent the observed (unobserved) nodes involved before (after) $t$. Given the behavioral dynamics of node $p_1$ represented by its survival rates, and the number of its offspring nodes that have involved before $t$, we can predict the cumulative number of its offspring nodes that involve in the cascade at any time $t'>t$. After conducting similar predictions on all the observed nodes, the cascading process after $t$ can be predicted by an additive function over all local predictions from behavioral dynamics.

More specifically, how to model behavioral dynamics and further predict cascading process based on continuous-time survival theory also entail many challenges. First, it is unclear what distribution form the behavioral dynamics follow. Although Exponential and Rayleigh distributions are commonly used to characterize the temporal scaling of pairwise interactions, behavioral dynamics in this paper are a reflection of collective behaviors and are proved to be inconsistent with these simple distributions in real data. Second, the parameters in survival models are difficult to interpret, which limits the generality of the learned model. Given the distribution form of data, the parameters of survival model can always be learned from real data in maximum likelihood manner. However, it is unsure what these parameters stands for and the learned model cannot be generalized to out-of-sample nodes (i.e. the nodes whose behavioral dynamic data is not included in the data). Third, the predictive models based on survival theory are computationally expensive due to the continuous-time characteristic, which makes them infeasible in real applications. Thus, we intend to design an effective and interpretable model for behavioral dynamics modeling and a scalable solution for cascading process prediction.

In particular, we conduct extensive statistical analysis on large scale real data and find that the behavioral dynamics cannot be well captured by simple distributions such as Exponential and Rayleigh distribution, but the general form of Exponential and Rayleigh, Weibull distribution, can well preserve the characteristics of behavioral dynamics. Also, we discover strong correlations between the parameters of a node's behavioral dynamics and its neighbor nodes behavioral features. Enlightened by these, we propose a NEtworked WEibull Regression (NEWER) model for parameter learning of behavioral dynamics. In addition to the maximum likelihood estimation term, we also assume the parameters of a node can be regressed by the behavioral features of its neighbor nodes and thus impose networked regularizers to improve the interpretability and generality of the model. Based on the behavioral dynamics, we further propose an additive model for cascading process prediction. To make it scalable, we propose an efficient sampling strategy for approximation with a theoretical guarantee.

\begin{figure}[htbg]
    \centering
    \includegraphics[width=0.5\textwidth]{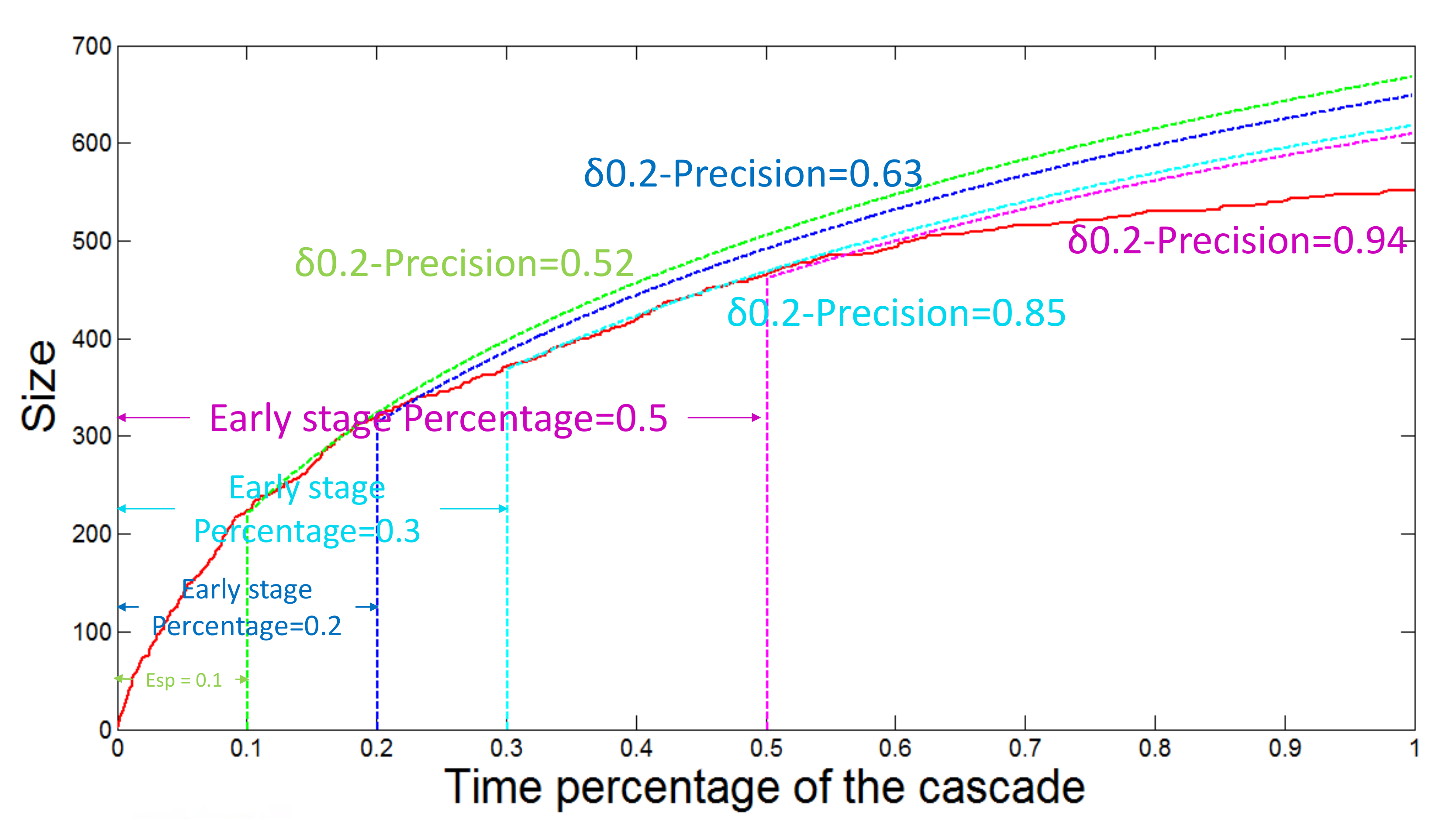}
    \vspace{-0.2in}

    \caption{Showcase of cascading process prediction for a real cascade. The red line represents the groundtruth cascading process. The others are prediction results based on different early stage information.}
    \label{fig:evo_cas}
    \vspace{-0.2in}
\end{figure}

We extensively evaluate the proposed method in a complete dataset from a population-level social network in China, including over 320 million users, 1.2 billion edges and 340 million cascades . In all the testing scenarios, the proposed method can significantly outperform other baseline methods. Figure \ref{fig:evo_cas} is a showcase of cascading process prediction by the proposed method. We show that by accurately modeling behavioral dynamics of social network users, we can predict the cascading process with a 2 hours leading time window, and get the average precision of $0.97$ if we restrict the error rate of the size to be $0.1$. Also, the accurate predictions of final cascade size, cascade outbreaking time are all implied in the predicted cascading process.

The main contributions of this paper are:

(1) Enlightened by the cascading size prediction works, we move one step forward to attempt cascading process prediction problem, which implies several vital problems such as cascade size prediction, outbreaking time prediction as well as evolving process prediction.

(2) We find out the common principles and patterns lying in behavioral dynamics and propose a novel Networked Weibull Regression model for behavioral dynamics modeling accordingly, which significantly improves the interpretability and generality of traditional survival models.

(3) We propose a novel method for predicting macro cascading process by aggregating micro behavioral dynamics, and propose a scalable solution to approximate the cascading process with a theoretical guarantee.

\vspace{-0.05in}
\section{RELATED WORK}
\noindent\textbf{Prediction on Cascades}. In recent years, many methods have been proposed to make prediction on cascades. Most of them focus on predicting the future size of a cascade, and the common way is to select vital nodes and place sensors on them. For example, Cohen {\em et al}. \cite{cohen2003efficient} focus on exploring the topological characteristics of the cascade. Cui et al. \cite{cui2013cascading} proposes to optimize the size prediction problem using dynamic information. Cheng et al. \cite{cheng2014can} introduces temporal feature into the problem and they predict the growing size of the cascade. Rather than attempt to predict the cascade size, we focus on predicting the cascading process which considers both time and volume information together.

\noindent\textbf{Survival Model}. Survival model is a method try to analysis things according to the time duration until one or more events happen. In recent years, researchers started modeling information diffusion using continuous models. Myers et al. \cite{myers2010convexity} proposed CONNIE to infer the diffusion network base on convex programming while leaving the transmission rate to be fixed, later on Rodriguez et al.\cite{rodriguez2011uncovering} proposed NETRATE which allowing the transmission rate to be different in different edges. Subsequently, Rodriguez et al. \cite{manuel2013icml} give an additive model and a multiplicative model to describe information propagation base on survival theory. Most of these works focus on discovering the rules and patterns to the edges in the social network and is hard to extend to make predictions for cascades since the correlation between transmission rates on edges is little. In contrast, our work focus more on predictive modeling by grouping correlated edges together so that we can make predictions for edges base on the information of other edges. 

\noindent\textbf{Influence Modeling and Maximization}. Influence modeling and maximization aims to evaluate users' importance in social networks. This is first proposed by Domingos et al. \cite{domingos2001mining} to select early starters to trigger a large cascade. Then Kempe et al. \cite{kempe2003maximizing} proposed Stochastic Cascade Model to formalize the problem and Chen et al. \cite{chen2009efficient} proposed a scalable solutions. Recently the approach was extended to adding opinion effect \cite{chen2011influence, gionis2013opinion} or time decay effects \cite{rodriguez2012influence} on the models. Our work is distinct from existing works in the following way: Rather than quantify the influence on nodes, we will predict the cascading process.

\vspace{-0.05in}
\section{Preliminaries}

This section presents the dataset information, discovered patterns and validated hypothesises to support the model design and solution.

\vspace{-0.05in}
\subsection{Dataset Description}\label{section:sec3.1}

\begin{figure}[htbg]
    \vspace{-0.105in}
    \centering
    \includegraphics[width=0.5\textwidth]{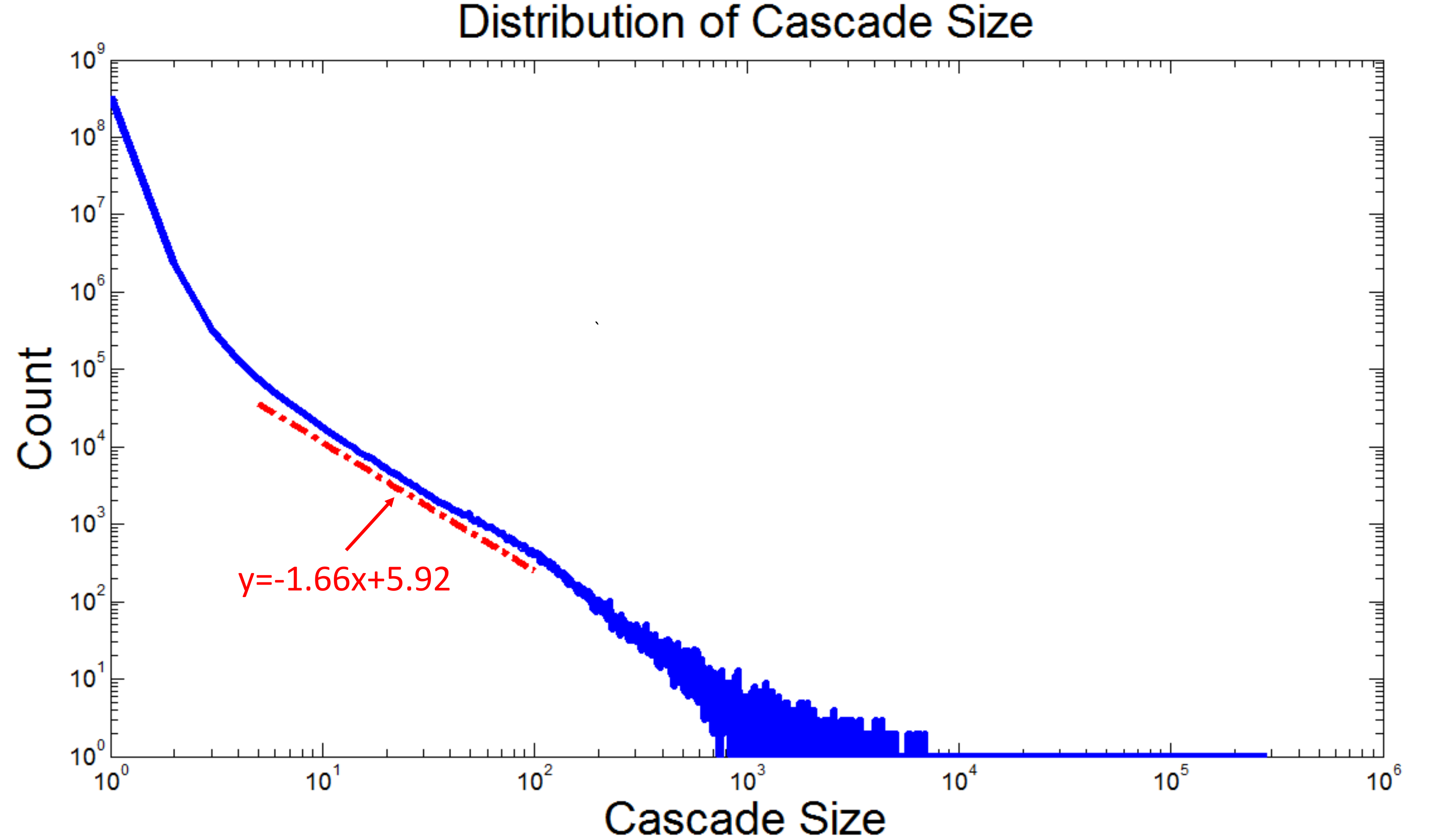}
    \vspace{-0.275in}

    \caption{Distribution of cascade size. The red straight line is the linear fitting result to the blue curve, showing the size distribution fits power-law.}
    \label{fig:cascade_size_distribution_figure}
    \vspace{-0.105in}
\end{figure}

The dataset in this paper is from Tencent Weibo, one of the largest Twitter-style websites in China. We collect all the cascades in 10 days generated between Nov 15th and Nov 25th in 2011. The dataset contains in total 320 million users with their social relations, 340 million cascades\footnote{Here the cascades are information cascades. When a user retweet/generate a post, several of his/her followers will further retweet the post and so on so forth to form a information cascade.} with their explicit cascading processes. The distribution of cascade size is shown in Figure \ref{fig:cascade_size_distribution_figure}. We can see that the cascade size follows Power-Law distribution, and the majority of cascades have very small size, which are not of interest for many applications. As the paper intends to predict cascading process, we filter out the cascades with the size of less than 5, and maintain the remaining 0.59 million cascades with obvious cascading process for statistical analysis and experiments.

\begin{table*}
    \begin{tabular}[c]{ c | p{3.5cm} | p{3.5cm} | p{3.5cm} | c }
        \hline \hline
        model & density function & survival function & hazard function & ks-static in Weibo \\
        \hline
        Exponential & $\lambda_ie^{-\lambda_it}$ & $e^{-\lambda_it}$ & $\lambda_i$ & $0.2741$ \\
        Power Law & $\frac{\alpha_i}{\delta}\left(\frac{t}{\delta}\right)^{-\alpha_i-1}$ & $\left(\frac{t}{\delta}\right)^{-\alpha_i}$ & $\frac{\alpha_i}{t}$ & $0.9893$ \\
        Rayleigh & $\alpha_ite^{-\alpha_i\frac{t^2}{2}}$ & $e^{-\alpha_i\frac{t^2}{2}}$ & $\alpha_it$ & $0.7842$ \\
        Weibull & $\frac{k_i}{\lambda_i}\left(\frac{t}{\lambda_i}\right)^{k_i-1}e^{-\left(\frac{t}{\lambda_i}\right)^{k_i}}$ & $e^{-\left(\frac{t}{\lambda_i}\right)^{k_i}}$ & $\frac{k_i}{\lambda_i}\left(\frac{t}{\lambda_i}\right)^{k_i-1}$ & \textbf{\color{red}{0.0738}} \\
        \hline
    \end{tabular}
    \vspace{-0.105in}
    \caption{Parametric Models}
    \label{table:survival_function_tables}
    \vspace{-0.15in}
\end{table*}

%

\vspace{-0.05in}
\subsection{Characteristics of Behavioral Dynamics}

\begin{figure}[htbg]
    \vspace{-0.05in}
    \centering
    \includegraphics[width=0.5\textwidth]{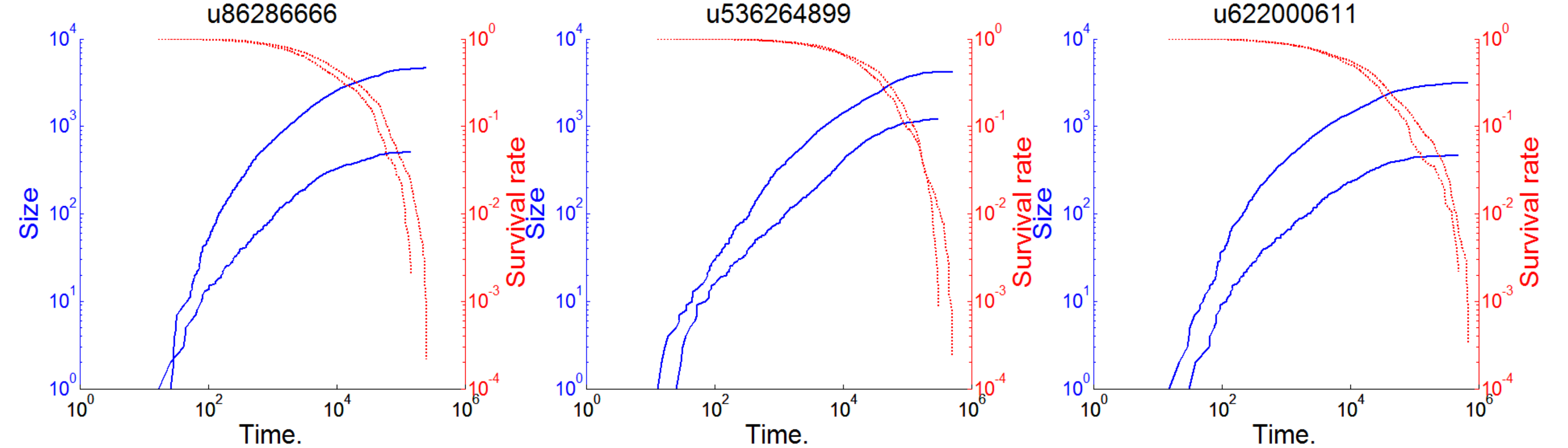}
    \vspace{-0.25in}

    \caption{The size growth curves and their corresponding survival function for 3 users.}
    \label{fig:multiple_user_different_cascades}
    \vspace{-0.105in}
\end{figure}

As mentioned before, behavioral dynamics play a central role in uncovering and predicting cascade processes. Here we investigate the characteristics of behavioral dynamics to enlighten the modeling of behavioral dynamics. By definition, the behavioral dynamics of a user capture the changing process of the cumulative number of his/her followers retweet a post after the user retweeting the post. Then the behavioral dynamics of a user can be straightforwardly represented by averaging the size growth curve of all subcascades that spread to the user and his/her followers. However, Figure \ref{fig:multiple_user_different_cascades} shows that the size growth curves vary significantly for different subcascades of the same user, which means that such a representation is not fit to characterize behavioral dynamics. Here we normalize the size growth process by the cascade final size and adopt survival function to describe the behavioral dynamics where the survival rate represents the percentage of nodes that has not been but will be infected. As shown in Figure \ref{fig:multiple_user_different_cascades}, a user's survival function is quite stable for different subcascades although their size growth patterns vary.


Then can we use the behavioral dynamics represented by survival function to predict the size growth curve of a subcascade? We provide positive answer with the assistance of early stage information. For example, if we know the subcascade size at an early time $t_0$, then the survival function can be straightforwardly transformed from percentage dimension into size dimension.

\vspace{-0.05in}
\subsection{Parametrize Behavioral Dynamics}

For the ease of computation and modeling, we need to parametrize the behavioral dynamics in our case. In state-of-the-art, Exponential and Rayleigh distributions are often used to describe the dynamics of user behaviors in different settings \cite{du2013uncover,gomez2013structure}. Here we testify these distribution hypothesis on our real data and find that these distributions cannot well capture both the shape and scale characteristics of behavioral dynamics. Thus, we turn to the general form of Exponential and Rayleigh distributions, the Weibull distribution \cite{pinder1978weibull}, and find it adequate for parametrizing behavioral dynamics. In order to quantify the effect of parametrization, we calculate KS-Statistic for the three candidate distributions as shown in Table \ref{table:survival_function_tables}. It displays that Weibull distribution performs much better than Exponential and Rayleigh distribution. The improvement is attributed to the high degree of freedom of Weibull distribution as it has two parameters $\lambda$ and $k$ to respectively control the scale and shape of the behavioral dynamics.

\vspace{-0.05in}
\subsection{Covariates of Behavioral Dynamics} \label{section:sec3.4}

\begin{figure}[htbg]
    \centering
    \includegraphics[width=0.5\textwidth]{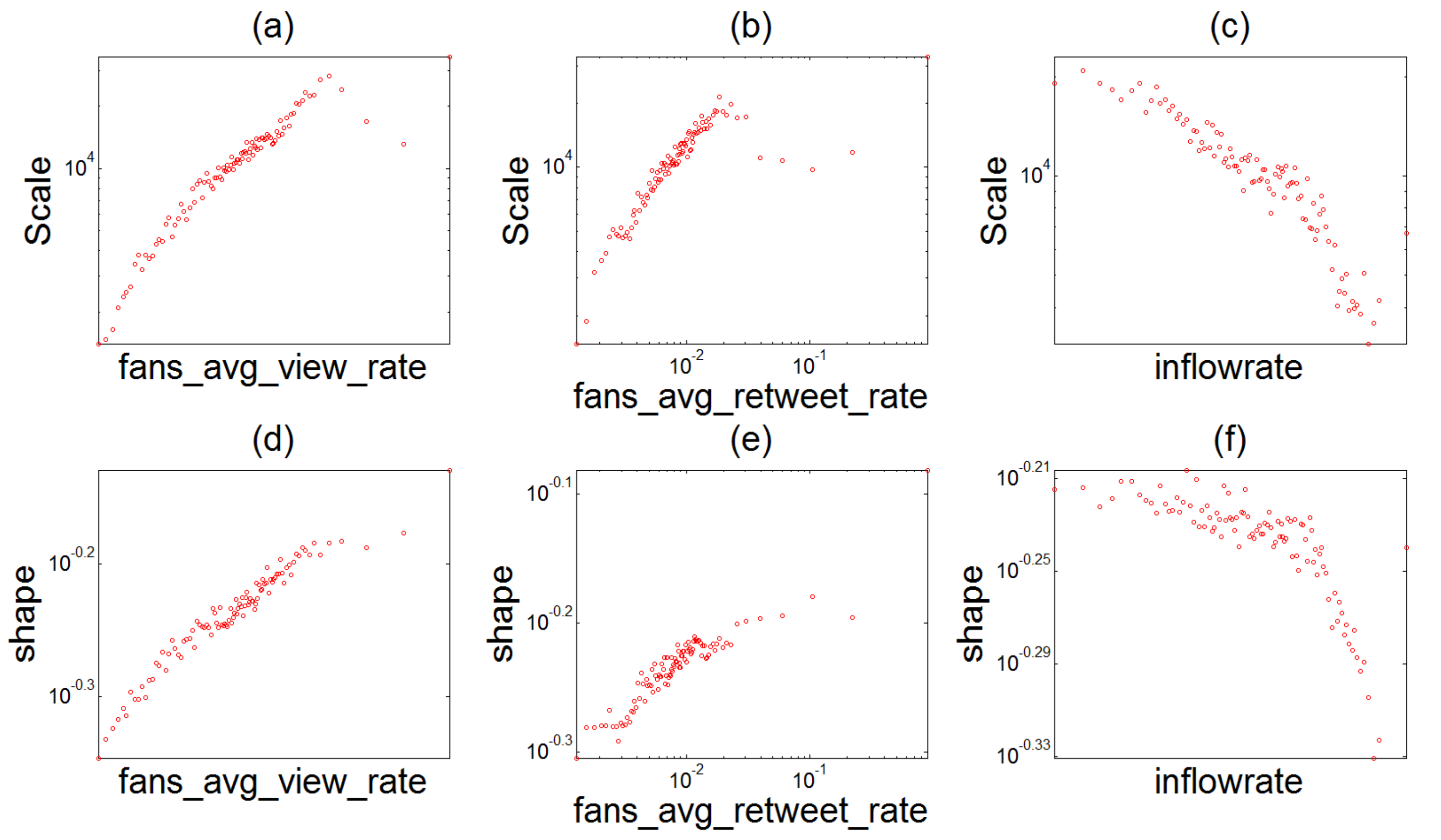}
    \vspace{-0.425in}

    \hspace{0.0002in}
    \caption{Correlations between the survival function parameters and the behavioral features}
    \label{fig:dsi_vs_parameters}
    \vspace{-0.23in}
\end{figure}

If subcascades for all users are sufficient, the parameters of behavioral dynamics can be directly learned from data. However this suffers from several drawbacks: (1) some users may have no or very sparse subcascade in training dataset, which makes these users' behavioral dynamics inaccurate or even unknown; (2) it is difficult to interpret the parameters directly learned from data, which prohibits us from getting insightful understanding on the behavioral dynamics. To address these, we investigate the covariates of behavioral dynamics here. As the behavioral dynamics of a user are to capture the collective responses of his/her followers, we assume the parameters of the user's behavioral dynamics should be correlated with the behavioral features of his/her followers (network neighbors). Hence, we extract a set of behavioral features for each user as listed in Table \ref{table:behavioral_feature_table}.\footnote{We think that follower with different retweet number will have different effects to the user, so we modify the weights on each term of $follower\_avg\_inflow\_rate$ and $follower\_avg\_retweet\_rate$. }  For each user with enough subcascades in our dataset, we learned their $\lambda$ and $k$ directly from data. And then, we calculate the correlations between the learned parameters and their followers' collective behavioral features. The examples given in Figure \ref{fig:dsi_vs_parameters} indicate obvious correlations between the learned parameters with these behavioral features. Therefore, we can use these behavioral features as covariates to regress the parameters of behavioral dynamics.

\begin{table}
    \vspace{0.05in}
    \small{
     \begin{tabular}[c] {l l}
     \hline
     Behavioral features \\
     \hline
     $inflow\_rate$ & the number of the posts user re-\\
     &ceived in a certain period. \\
     \hline
     $outflow\_rate$ & the number of the posts user sent\\
     & in a certain period. \\
     \hline
     & average inflow rate of fans to the\\
     $follower\_avg-$ & user, or $\frac{\sum_{i}{retweet(i)\cdot in\_flow(i)}}{\sum_{i}{retweet(i)}}$\\
     $\_inflow\_rate$& where i is the fans to the user(and\\
     & the same as following).\\
     \hline
     $follower\_avg-$ & average retweet rate of fans to the\\
     $\_retweet\_rate$ & user, or $\frac{\sum_{i}{retweet(i)\cdot retweet\_rate(i)}}{\sum_{i}{retweet(i)}}$.\\
     \hline
     Structural features \\
     \hline
     $follower\_number$ & number of the followers to the user.\\
     \hline
     $follow\_number$ & number of users this user follows.\\
     \hline
     \end{tabular}
    }
    \vspace{-0.1in}

    \caption{Behavioral features for users.}
    \label{table:behavioral_feature_table}
    \vspace{-0.155in}

\end{table}

%

\vspace{-0.05in}
\subsection{From Behavioral Dynamics to Cascades}

After validating that the behavioral dynamics can potentially be accurately modeled and predicted, the key problem is whether we can derive the macro cascading process from micro behavioral dynamics. Intuitively, the cascading process cannot be perfectly predicted at early stage by behavioral dynamics. Given any time $t$, we can only use the behavioral dynamics of the users that involved before $t$ to predict the cascading process after $t$. Consequently, the prediction coverage is restricted to all the followers of these users, while the users beyond this scope are neglected. These uncovered users may potentially affect the performance of cascading process prediction.

Fortunately, we observe two interesting phenomenons in real data.

\begin{figure}[htbg]
    \vspace{-0.105in}
    \centering
    \includegraphics[width=0.5\textwidth]{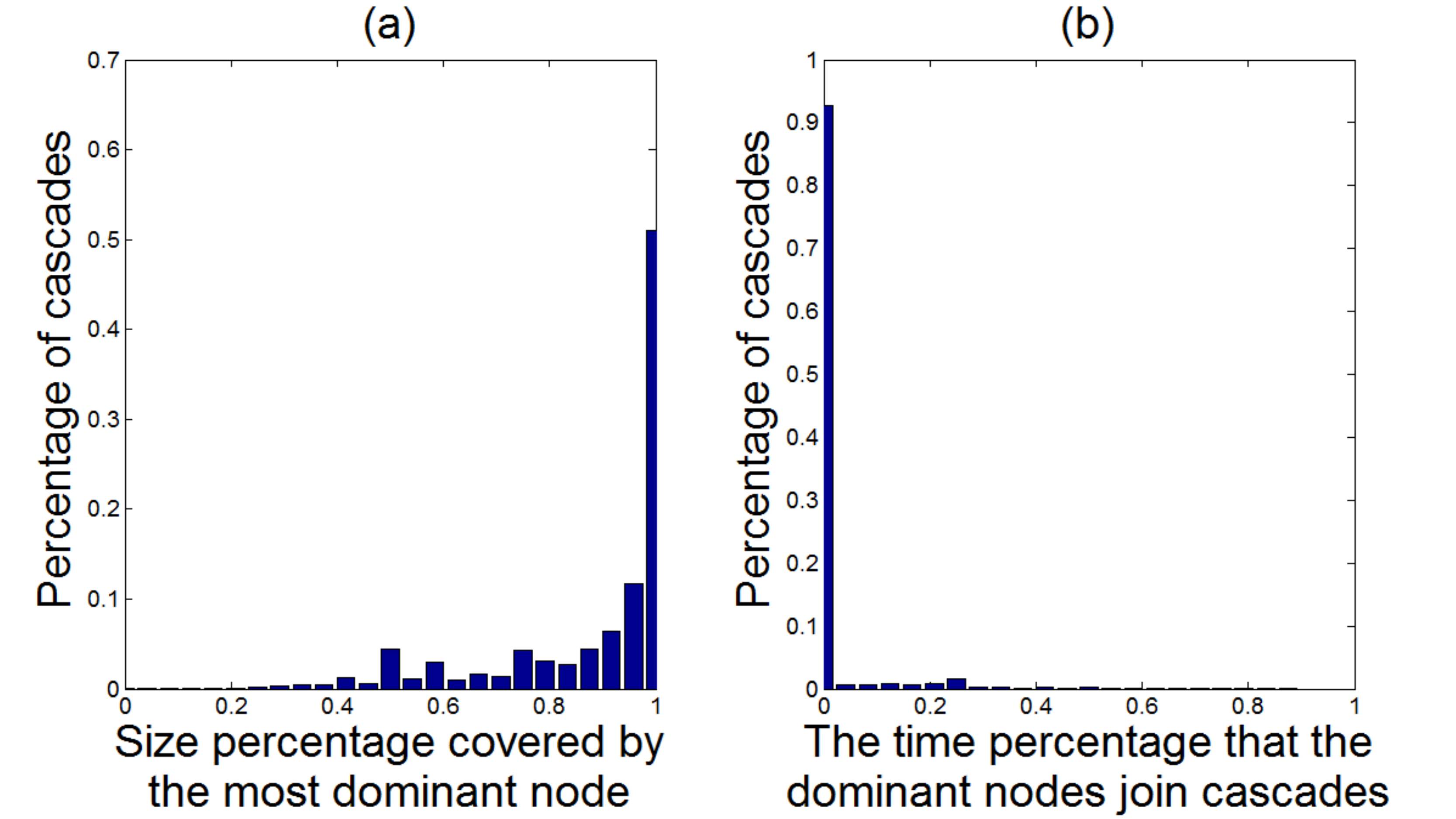}
    \vspace{-0.275in}

    \caption{Minor dominance and early stage dominance in information cascades.}
    \label{fig:dominant_nodes}
    \vspace{-0.105in}
\end{figure}

\textbf{Minor dominance}. Although each user has behavioral dynamics, the behavioral dynamics of different users make significantly different contributions to the cascading process. It is intuitive that the behavioral dynamics of an active user with 1 million followers contribute much more than that of an inactive user with 5 followers. The data also coincides with our intuition. According to Figure \ref{fig:dominant_nodes} (a), it can be observed that a very small number of nodes whose behavioral dynamics dominate the cascading process underpin the idea of just using the behavioral dynamics of these dominant nodes for cascading process prediction.

\textbf{Early stage dominance}. Enlightened by the minor dominance phenomenon, we further ask whether the dominant nodes are prone to join cascades in early stage. Here, Figure \ref{fig:dominant_nodes} (b) depicts the time distribution of these dominant nodes joining in cascades, and we can see that most of these nodes in actual join cascades in the very early stage.

%

Taking these two phenomena into account together, it is safe to design a model exploiting the behavioral dynamics of infected nodes in early stage to predict the cascading process.

\section{METHODOLOGY}

This section introduces the NEtworked WEibull Regression (NEWER) and cascade prediction methods in detail.

\vspace{-0.05in}
\subsection{Problem Statement}

Given a network $G=\langle U, A\rangle$, where $U$ is a collection of nodes and $A$ is the set of pairwise directed/undirected relationships. An event (e.g., tweet) can be originated from one node and spread (e.g., by retweeting) to its neighboring nodes. A cascade is typically formed by repeating this process. Therefore, a cascade can be represented by a set of nodes $C=\{u_1,u_2,...u_m\}$, where $u_1$ is the root node. 
In a cascade, each node will get infected by the event only once, so it is tree-structured. For every node  $u_{i}$ in the cascade, we denote its parent node as $rp(u_i)$. The time stamp that $u_i$ gets infected is $t(u_i)$, and $t(u_{i})\leq t(u_{i+1})$. Then the partial cascade before time $t$ is denoted by $C_{t}=\{u_i|t(u_i)\leq t\}$, and its size $size(C_{t})=|C_{t}|$ where $|.|$ is the cardinality of a set. Then the cascade prediction problem can be defined as: \\[-.8em]

\noindent\textbf{Cascade Prediction}: Given the early stage of a cascade $C_{t}$, predict the cascade size $size(C_{t'})$ with $t'\geq t$.

\vspace{-0.05in}
\subsection{Survival Analysis}


Survival analysis is a branch of statistics that deals with analysis of time duration until one or more events happen, such as death in biological organisms and failure in mechanical systems \cite{miller2011survival}. It is a useful technique for cascade prediction. More concretely, let $\tau_0$ be a non-negative continuous random variable representing the waiting time until the occurrence of an event with probability density funtion $f(t)$, the \textbf{survival function}
\begin{equation}
	S(t)=Pr\{\tau_0 \geq t\}=\int_{t}^{\infty}{f(t)}
\end{equation}
encodes the probability that the event occurs after $t$, the \textbf{hazard rate} is defined as the event rate at time $t$ conditional on survival until time t or later ($\tau_0\ge t$), i.e.,
\begin{equation}
	\lambda(t)=\lim_{dt \rightarrow 0}\frac{Pr(t\leq \tau_0<t+dt | \tau_0\geq t)}{dt}=\frac{f(t)}{S(t)}
\end{equation}
\vspace{-0.125in}

$S(t)$ and $\lambda(t)$ are the two core quantities in survival analysis.

\vspace{-0.05in}
\subsection{NEtworked WEibull Regression Model}

The Weibull distribution is commonly used in survival analysis.
 In network scenario, if we think the time that an event (e.g., retweet) happened on a node as a survival process, we can fit a Weibull distribution to the survival time of node $i$, then its corresponding density, survival and hazard functions
    \begin{align}
        f_i(t) &= \frac{k_i}{\lambda_i}\left(\frac{t}{\lambda_i}\right)^{k_i-1}exp^{-\left(\frac{t}{\lambda_i}\right)^{k_i}}\\
        S_i(t) &= exp^{-\left(\frac{t}{\lambda_i}\right)^{k_i}}\\
        h_i(t) &= \frac{k_i}{\lambda_i}\left(\frac{t}{\lambda_i}\right)^{k_i-1}
    \end{align}
where $t>0$ is the average event happening time to node $i$, $\lambda_i>0$ and $k_i>0$ is the scale and shape parameter of the Weibull distribution. In the following we will assume the network nodes are users and the event is retweeting.\\[-.8em]

\noindent\textbf{Likelihood of retweeting dynamics}. Supposing there are $N$ users in total, $T_i$ is a set of $m_i$ time stamps and each element $T_{i,j}$ indicates the $j$-th retweet time stamp to the post of the $i$-th user. We sort those time stamps out in increasing order so that $T_{i,j+1}\geqslant T_{i,j}$. We assume $T_{i, j} \geq 1$ and $T_{i,m_i} > 1$. Then the likelihood of the event data can be written as follows:
\vspace{-0.05in}
\begin{align}
		L(\lambda, k) &= \prod_{i=1}^N\prod_{j=1}^{m_i}\left(h_i(T_{i,j}) \cdot S_i(T_{i,j})\right) \nonumber\\
				& = \prod_{i=1}^N\prod_{j=1}^{m_i}{\left(k_i \cdot T_{i,j}^{k_i-1} \cdot \lambda_i^{-k_i} \cdot e^{-T_{i,j}^{k_i} \cdot \lambda_i^{-k_i}}\right)} \\
		\log{L(\lambda, k)} &= \sum_{i=1}^N{l_i(\lambda_i,k_i)} \label{eq:loglikelihoodfunction}
\end{align}
\vspace{-0.05in}
where $l_i(\lambda_i,k_i)=m_i\log{k_i}+(k_i-1)\sum_{j=1}^{m_i}{\log{T_{i,j}}}-$\\$m_ik_i\log{\lambda_i}-\lambda_i^{-k_i}\sum_{j=1}^{m_i}{T_{i,j}^{k_i}}$.

As discovered in section \ref{section:sec3.4}, the survival characteristics of the user is correlated with the behavioral features of him/her. Then we can parametrize those parameters in the personalized Weibull distributions using those behavioral features. More formally, let $x_i$ be a $r$ dimensional feature vector for user $i$, we parameterize $\lambda_i$ and $k_i$ with the following linear function:
	\begin{align}
		\log\lambda_i=\log {x_i}*\beta \label{eq:lxb}\\
		\log{k_i}=\log{x_i}*\gamma \label{eq:kxg}
	\end{align}
where $\beta$ and $\gamma$ are $r$-dimensional parameter vector for $\lambda$ and $k$. We attempt to find the scale and shape parameter of every user so that the likelihood of the observed data is maximized, at the same time we can also get the parameter vectors for out-of-sample extensions.

We use the Equation (\ref{eq:lxb}) and (\ref{eq:kxg}) to replace $\lambda_i$ and $k_i$ in the log likelihood function Equation (\ref{eq:loglikelihoodfunction}) to solve the parameters. To further enhance the interpretability, we also add $\ell$1 sparsity regularizers on $\beta$ and $\gamma$ respectively to enforce model sparsity. Combining everything together, we can obtain the {\em NEtworked WEibull Regression} (NEWER) formulation which aims to minimize the following objective:
	\begin{align}
        F(\lambda, k, \beta, \gamma) &= G_1(\lambda, k) + \mu G_2(\beta, \lambda) + \eta G_3(\gamma, k)\label{eq:objectivefunction}\\
		G_1(\lambda, k) &= -\log{L(\lambda, k)}\\
		G_2(\lambda, \beta) &= \frac{1}{2N}\left\|\log\lambda-\log X\cdot\beta\right\|^2+\alpha_\beta\left\|\beta\right\|_1\\
		G_3(k, \gamma) &= \frac{1}{2N}\left\|\log k-\log X\cdot\gamma\right\|^2+\alpha_\gamma\left\|\gamma\right\|_1
	\end{align}

\noindent\textbf{Optimization}. To minimize $F(\lambda, k, \beta, \gamma)$  in Equation (\ref{eq:objectivefunction}), we first prove that the function is lower bounded. We have the following theorem.\\[-.8em]

\vspace{-0.05in}
\begin{theorem}
    $F(\lambda, k, \beta, \gamma)$  has global minimum.
\end{theorem}
\begin{proof}
    See the appendix.
\end{proof}
\vspace{-0.05in}

With this theorem, the following coordinate descent strategy can be used to solve the problem with guaranteed convergence. At each iteration, we solve the problem with one group of variables with others fixed.
    \vspace{-0.155in}

	\begin{eqnarray}
		\left\{
			\begin{aligned}
				For~it & = 1, \ldots, it_{max}\\
					& \lambda^{[it+1]}={argmin}_{\lambda}{F(\lambda, k^{[it]}, \beta^{[it]}, \gamma^{[it]})}\\
					& k^{[it+1]}={argmin}_{k}{F(\lambda^{[it+1]}, k, \beta^{[it]}, \gamma^{[it]})}\\
					& \beta^{[it+1]}={argmin}_{\beta}{F(\lambda^{[it+1]}, k^{[it+1]}, \beta, \gamma^{[it]})}\\
					& \gamma^{[it+1]}={argmin}_{\gamma}{F(\lambda^{[it+1]}, k^{[it+1]}, \beta^{[it+1]}, \gamma)}\\
			\end{aligned}
		\right.
	\end{eqnarray}

For solving the subproblem with respect to $\lambda$ or $k$, we use Newton's Method. For subproblem with respect to $\beta$ and $\gamma$, we use standard LASSO solver \cite{tibshirani1996regression}.


\vspace{-0.05in}
\subsection{Efficient cascading process prediction} \label{section:sec4.2}

%
%
%

It should be born in mind that cascading prediction is intended to perform early prediction of its size at any later time. In the following we will present two models to achieve this goal.

\vspace{-0.05in}
\subsubsection{Basic Model}

The entire flow of the basic model we proposed is illustrated in Algorithm \ref{alg: Basemodel}:

\vspace{-0.1in}
\begin{algorithm}[htb]
	\caption{Basic Model}
	\label{alg: Basemodel}{\small
	\begin{algorithmic}[1]
		\REQUIRE ~~\\
			Set of users $U$ involved in the cascade $C$ before time $t_{limit}$, survival functions of users $S_{u_j}(t)$, predicting time $t_e$;
		\ENSURE ~~\\
			Size of cascade $size\left(C_{t_e}\right)$;
		\FORALL {user $u_i \in U$}
			\STATE creates a subcascade process with $replynum(u_i) = 0$
			\IF {$u_i$ is not root node}
				\STATE $replynum(rp(u_i)) = replynum(rp(u_i)) + 1$\
			\ENDIF
		\ENDFOR
		\STATE $sum = 1$
		\FORALL {user $u_i \in U$}
            \STATE $deathrate(u_i)=\max\left(1-S_{u_i}(t_{limit}-t(u_i)), \frac{1}{|V|}\right)$ \label{code:recentStart1}\
            \STATE $fdrate(u_i)=\max\left(1-S_{u_i}(t_e-t(u_i)), \frac{1}{|V|}\right)$ \label{code:recentStart2}\
			\STATE $sum = sum + \frac{replynum(u_i)\cdot fdrate(u_i)}{deathrate(u_i)}$\
		\ENDFOR
		\RETURN $size\left(C_{t_e}\right)=sum$
	\end{algorithmic}}
\end{algorithm}
\vspace{-0.1in}

When a new node $u_i$ is added into the cascade at $t(u_i)$, the algorithm will launch a process to estimate the final size of the subcascade that $u_i$ will generate, with temporal size counter $replynum(u_i)$ and survival function $S_{u_i}(t)$ starting at $t(u_i)$. If $u_i$ is involved by others, the algorithm also increases the temporal size of the retweet set of its parent $rp(u_i)$ by one.

After all the information before the deadline is collected, the result will be finalized by aggregating all the value estimated by every subcascade process. 
Since the post number is at most $|V|$ (all nodes in the network are involved into the cascade), the value of death rate $deathrate(u_i)$  and final death rate $fdrate(u_i)$ (complement to their survival rates) at line \ref{code:recentStart1} and line \ref{code:recentStart2} is set to be $1/|V|$ when it is lower than $1/|V|$.


\noindent\textbf{Complexity Analysis}. Only constant time operations is involved in the two for-loops. Therefore, the complexity of the algorithm is $O(n)$ where $n$ is the number of users in the cascade.
\vspace{-0.05in}

\subsubsection{Sampling Model}

Although the basic model solves the estimation problem, real applications often need to estimate the cascade size dynamically so that the changes can be monitored. 

%
%
%

To make the algorithm scalable, the number of recalculations should be limited, while the estimated value of size should fall into an acceptable error scope. We can utilize the following two facts to make the estimation process more efficient: (1) For a subcascade generated by $u_i$, the estimation of the size will always be zero if there is no user involved into it, which means we can ignore the calculation.
(2) If we do not re-estimate the final number of a subcascade (when there is no new user involved into it), the temporal size counter $replynum(u_i)$ and final death rate $edrate(u_i)$ will not change but the death rate $deathrate_{u_i}(t)$ will increase over time. Supposing the previous time stamp of the subcascade set estimation is $t_0$, it will cause a relative error rate of $\frac{deathrate_{u_i}(t_1)}{deathrate_{u_i}(t_0)}-1$ at $t_1$. Hence, the relative error rate will be at most $\epsilon$ if we re-estimate the final number of the subcascade at $S_{u}^{-1}{(1-(1+\epsilon)\cdot(deathrate_{u_i}(t_0)))}$. By exploring those two tricks, we propose a sampling model shown in Algorithm \ref{alg: Improvedmodel}:

\begin{algorithm}[htb]
	\caption{ Sampling Model }
	\label{alg: Improvedmodel}{\small
	\begin{algorithmic}[1]
		\REQUIRE ~~\\
			survival functions of users $S_{u_j}(t)$, and set of users $U$ in one cascade $C$(given dynamically);
		\ENSURE ~~\\
			for every size prediction request to $t_e$ at $t_0$, output $size\left(C_{t_e}\right)$;
		\STATE $sum = 0$;\
		\WHILE {request = model.acceptRequest}
            \SWITCH{request.type}
    			\CASE {APPROXIMATION}
    				\RETURN $size\left(C_{t_e}\right) = sum$
    			\ENDCASE
    			\CASE {INVOLVED$\_$USER}
                    \STATE $u_i$=request.user, $t_0$=request.time \
    				\STATE creates a subcascade process: \\
                        $~~~t(u_i)=t_0$, $app(u_i)=0$, $replynum(u_i) = 0$,\\
                        $~~~fdrate(u_i)=\max\left(\frac{1}{|V|},1-S_{rp(u_i)}(t_e-t_0)\right)$;\
    				\IF {$u_i$ is root node}
    					\STATE $sum = 1$;
    				\ELSE
    					\STATE $t_{rep} = t_0-t\left(rp(u_i)\right)$;\
    						\STATE $replynum(rp(u_i))=replynum(rp(u_i))+1$;\
    						\STATE $sum = sum - app(rp(u_i))$;\
    					\STATE $deathrate(rp(u_i)) = \max\left(\frac{1}{|V|},1-S_{rp(u_i)}(t_{rep})\right)$;\
    					\STATE $app(rp(u_i)) = \frac{replynum(rp(u_i))\cdot fdrate(rp(u_i))}{deathrate(rp(u_i))}$;\
    					\STATE $sum = sum + app(rp(u_i))$;\
    					\STATE
                                $t_{new} = S_{rp(u_i)}^{-1}{(1-(1+\epsilon)\cdot deathrate(rp(u_i)))}$\\
                                $\hspace{1.25cm} + t(rp(u_i))$;\
                        \STATE sendRequest(THRESHOLD$\_$CHANGE,$rp(u_i)$,$t_{new}$);\
    				\ENDIF
    			\ENDCASE
    			\CASE {THRESHOLD$\_$CHANGE} 
                    \STATE $u_i=request.user$, $t_0$=request.time \
    				\STATE $sum = sum - app(u_i)$;\
    				\STATE $deathrate(u_i) = \max\left(\frac{1}{|V|}, 1-S_{u_i}(t_0-t(u_i))\right)$;\
    				\STATE $t_{new} = S_{u_i}^{-1}{(1-(1+\epsilon)\cdot deathrate(u_i))} + t(u_i)$;\
                    \STATE sendRequest(THRESHOLD$\_$CHANGE,$u_i$,$t_{new}$);\
    				\STATE $app(u_i) = \frac{replynum(u_i)\cdot fdrate(u_i)}{deathrate(u_i)}$;\
    				\STATE $sum = sum + app(u_i)$;\
    			\ENDCASE
            \ENDSWITCH
		\ENDWHILE
	\end{algorithmic}}
\end{algorithm}

\noindent\textbf{Complexity Analysis.} The following theorem analyzes the complexity of Algorithm 2.
\vspace{-0.05in}
\begin{theorem}
    With an overall $O(n~log_{1+\epsilon}(|V|))$ counting to estimate the number of subcascades, the sampling model can approximate the final size of the whole cascade at any time with an relative error rate of at most $\epsilon$.
\end{theorem}
\vspace{-0.105in}
\begin{proof}
	For each approximation request, we only need to report the number directly; for every new subcascade, the initially operation number is also constant, and we need to do at most $O(log_{1+\epsilon}(|V|))$ times threshold adjustment for subcascade which has users involved in, since the lowerbound of $deathrate$ is $\frac{1}{|V|}$ and the upperbound is $1$(all the people are involved in the cascade). Above all, the final complexity is $O(t)+O(n)+O(n~log_{1+\epsilon}(|V|))=O(t+n~log_{1+\epsilon}(|V|))$ for each cascade (with $n$ users and t requests). If we put this algorithm into an online environment, the complexity will be $O(T+N~ log_{1+\epsilon}(|V|)) \sim O(T)$ for all the cascades with $N$ Users in total\footnote{It will be counted multiple times if a specific user involves in multiple cascades} (we see $log_{1+\epsilon}(|V|)$ as a constant with respect to T and $N\sim T$ as the number of users involved in cascades increases over time).
\end{proof}
\vspace{-0.05in}

With this model, for cascade final size prediction, we just need to set the prediction time $t_e$ to be infinite so that the $deathrate$ of all subcascades will be 1. For outbreak time prediction, we can make a binary search with respect to time $t_e$, checking whether the cascade size will be more or less than the size number at $t_{mid}$ and make the decision eachtime.We can also solve the process prediction problem by asking the outbreak time of size restricted from the current size to the final size of the cascade one by one. In all cases, the sampling model will be n times faster than the base model.

\vspace{-0.05in}

\section{EXPERIMENTS}

In order to evaluate the performances and fully demonstrate the advantages of the proposed method, we conduct a series of experiments on the dataset introduced in Section \ref{section:sec3.1}. The results of multiple tasks are reported, including cascade size prediction, outbreak time prediction and cascading process prediction. Also,it will address more insights about the proposed method.



\vspace{-0.05in}
\subsection{Baselines and Evaluation Metrics}

Since we are the first to investigate cascading process prediction problem, no previous models can be adopted as direct baselines. Here, we implemented the following methods which can be potentially applied into our targeted problem as baselines:
    \vspace{-0.055in}
\begin{itemize}
    \item Cox Proportional Hazard Regression Model (Cox) : This model assumes that the behavioral dynamics of all users have different scale parameters while sharing the same shape parameter. We use the same covariates as in our model and find the optimal scale parameters for all users and the shared shape parameter. We implement it as in \cite{cox1972regression}.
    \vspace{-0.055in}
    \item Exponential/Rayleigh Proportional Hazard Regression Model (Exponential/Rayleigh): Since the shape parameters of both Exponential and Rayleigh distributions are fixed values (1 for Exponential distribution and 2 for Rayleigh distribution), they are two special cases of Cox model.
    \vspace{-0.055in}
    \item Log-linear Regression Model (Log-linear): We refer to \cite{cheng2014can} which extracted 4 classes features to characterize cascades, including node features, structural features of cascades, temporal features and content features. In our case, we ignore the content features which are not covered in our dataset and also reported by \cite{cheng2014can} to be unimportant for cascade prediction. Then we use log-linear regression model to predict the cascade size.
    \vspace{-0.125in}
\end{itemize}
    \vspace{-0.055in}

It is noted that Log-linear can only predict cascade size but not for time-related prediction, while Cox, Exponential and Rayleigh models are applied to all prediction tasks. Also, the goal of Cox, Exponential and Rayleigh models are to elucidate the behavioral dynamics. After that, we use the same cascade prediction model as in our method to conduct cascade-level predictions.


For each cascade, our dataset includes its complete cascading process as the groundtruth. Next, we use the following metrics to evaluate the performances:

\begin{itemize}
    \vspace{-0.055in}
    \item Root Mean Square Log Error ($RMSLE$): In Power-Law distributed data, it is not reasonable to use standard RMSE to evaluate the prediction accuracy. For example, for a cascade with the groundtruth size of 1000, it is significantly different to predict its size to be 2000 or 0, but they have the same RMSE. Thus, we first calculate the logarithmic results for both the groundtruth and predicted value, then calculate RMSE on the logarithmic results to evaluate the accuracy of the proposed method and baselines.
    \vspace{-0.055in}
    \item Precision with $\sigma$-Tolerance ($\delta\sigma$-Precision): In real applications, a small deviation from the groundtruth value is often acceptable. In our case, we regard the predicted value within the range of groundtruth$(1\pm\sigma)$ as a correct prediction, and the resulted precision is $\delta\sigma$-Precision.
    \vspace{-0.055in}
\end{itemize}

For parameter setting, there are 4 parameters in our method, including $\mu, \eta, \alpha_{\beta}$ and $\alpha_{\gamma}$. We tune these parameters by grid searching, and the optimal parameters used in our experiments are $\mu=10, \eta=10, \alpha_{\beta}=6*10^{-5}, \alpha_{\gamma}=8*10^{-6}$.


\vspace{-0.05in}
\subsection{Cascade Size Prediction}
\begin{figure}[htbg]
    \centering
    \vspace{-0.075in}
    \includegraphics[width=0.5\textwidth]{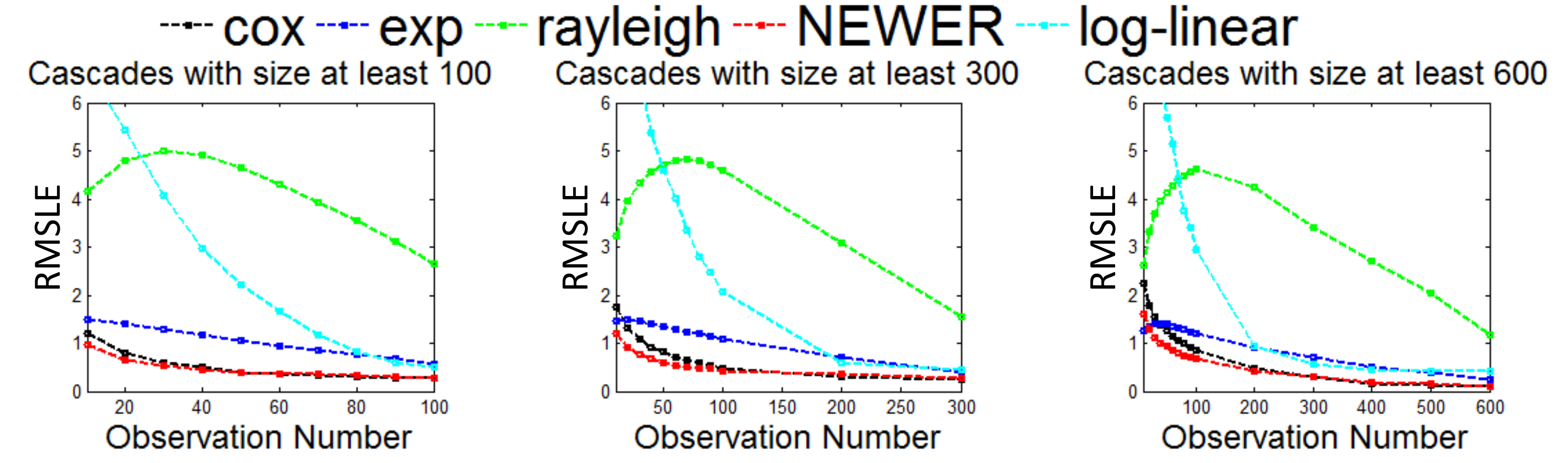}
    \vspace{-0.425in}

    \hspace{0.0002in}
    \caption{\textbf{$RMSLE$ results of different methods with different number of observed nodes in cascades.}}
    \label{fig:exp_rmse}
    \vspace{-0.125in}
\end{figure}

We randomly separate the cascades into 10 folds, and conduct a 10-fold cross validation by using 9 of them as training data and the other one as testing data. For cascades with size over $k$, we use the first $s(s<k)$ nodes as observed data, and the target is to predict the final cascade sizes.

The prediction performances of all the methods are shown in Figure \ref{fig:exp_rmse}. It can be seen that the proposed method NEWER significantly outperforms other baselines in $RMSLE$ value in different sized datasets. The baselines that has the closest performance with NEWER is the Cox model. We can see that the margins of improvement from Cox to NEWER are more obvious in the dataset with larger $k$. In a certain dataset, the margins are more evident with smaller $s$. These results demonstrate the significant advantage of NEWER in predicting large cascades in very early stage.

Comparatively, the Log-linear method does not achieve satisfactory results in this task. The main reason is that the coefficients in the Log-linear model are highly biased towards the dominant number of small-sized cascades, which is also argued by \cite{cheng2014can}. In our method, we successfully overcome this bias by shifting from macro cascade level features to micro behavioral dynamics. The substantial gain achieved by all behavioral dynamics based methods (including NEWER, Cox, Exponential and Rayleigh) exemplifies the importance of this micro mechanism for cascade prediction.

In order to demonstrate the efficiency of the proposed method, we also evaluate the computational cost of NEWER and Sampling-NEWER in the computational environment with 3.4GHZ Quad Core Intel i7-3770 and 16GB memory. We track the process of all cascades. The base cascade prediction model (Base) re-predicts the final size at every time points (in second), while the sampling-based cascade prediction model (Sampling) re-predicts the final size only when the observed cascade sizes increase. As shown in Table \ref{table:calculation_number_table}, the Sampling model (with a 10 percent performance degradation tolerance) is much more efficient than Base model by almost 5 magnitudes. According to Section \ref{section:sec4.2}, it is guaranteed that the Sampling method can also improve with similar magnitudes than the Base model in cascading process prediction task. So we omit these results for brevity.

\begin{table}
    \small{
     \begin{tabular}[c] {l | c | c | c}
     \hline
     Method & Base & Improved & Directed \\
      & Model & Model ($\delta=0.1$) & Learning Method \\
     \hline
     Size $\geq 20$ & $8.47*{10}^{5}s$ & $10.73s$ & $899s$ \\
     \hline
     Size $\geq 50$ & $7.61*{10}^{5}s$ & $8.62s$ & $899s$ \\
     \hline
     Size $\geq 100$ & $6.65*{10}^{5}s$ & $7.09s$ & $898s$\\
     \hline
     Size $\geq 500$ & $4.35*{10}^{5}s$ & $4.33s$ & $891s$\\
     \hline
     Size $\geq 1000$ & $3.4*{10}^{5}s$ & $3.30s$ & $881s$\\
     \hline
     \end{tabular}
    }
    \vspace{-0.15in}

    \caption{\textbf{Running time for different methods in different dataset under a server with 3.4GHZ Quad Core Intel i7-3770 CPU and 16GB memory.}}
    \label{table:calculation_number_table}
    \vspace{-0.205in}

\end{table}

\vspace{-0.05in}
\subsection{Outbreak Time Prediction}

\begin{figure}[htbg]
    \vspace{-0.105in}
    \centering
    \includegraphics[width=0.5\textwidth]{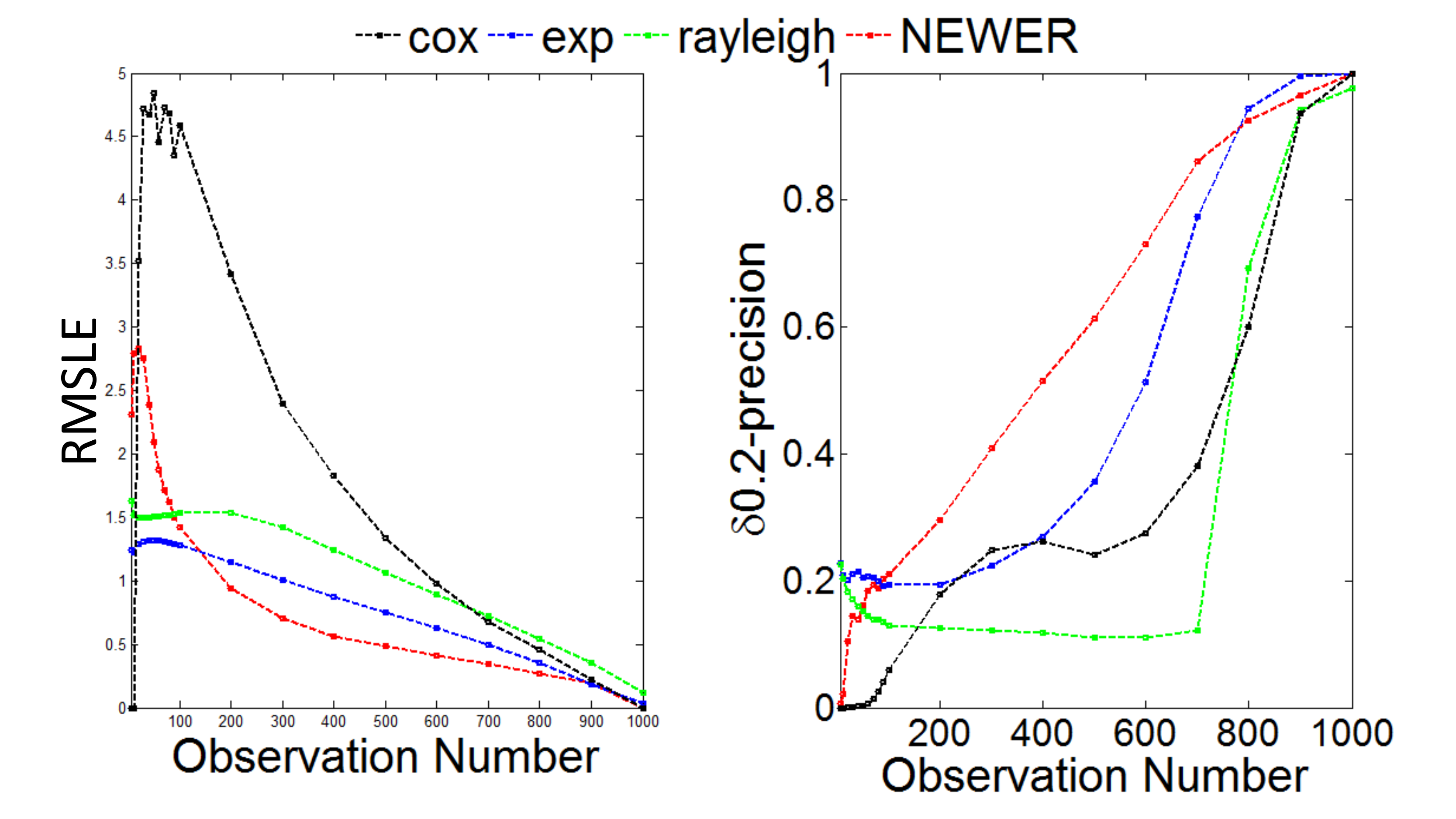}
    \vspace{-0.375in}

    \caption{Outbreak time prediction results of different methods with different number of observed nodes in cascades.}
    \label{fig:outbreak_time_prediction}
    \vspace{-0.205in}
\end{figure}

Another interesting problem is to predict when a cascading outbreak will happen. For example, in the early stage of a cascade, can we predict when the cascade reaches a specific size? Without loss of generality, we set the outbreak size threshold to be $1000$. We evaluate the prediction performance with different number of observed nodes in the cascades. As shown in Figure \ref{fig:outbreak_time_prediction}, the NEWER model get the best performances in both $RMSLE$ and $\delta\sigma$-Precision metrics. Although Exponential and Rayleigh models report better results than NEWER in very early stage (less than 50 observation nodes), the improvements of their performances with increasing number of observed nodes are not as significant as NEWER.

\subsection{Cascading Process Prediction}

\begin{figure}[htbg]
    \vspace{-0.135in}
    \centering
    \includegraphics[width=0.5\textwidth]{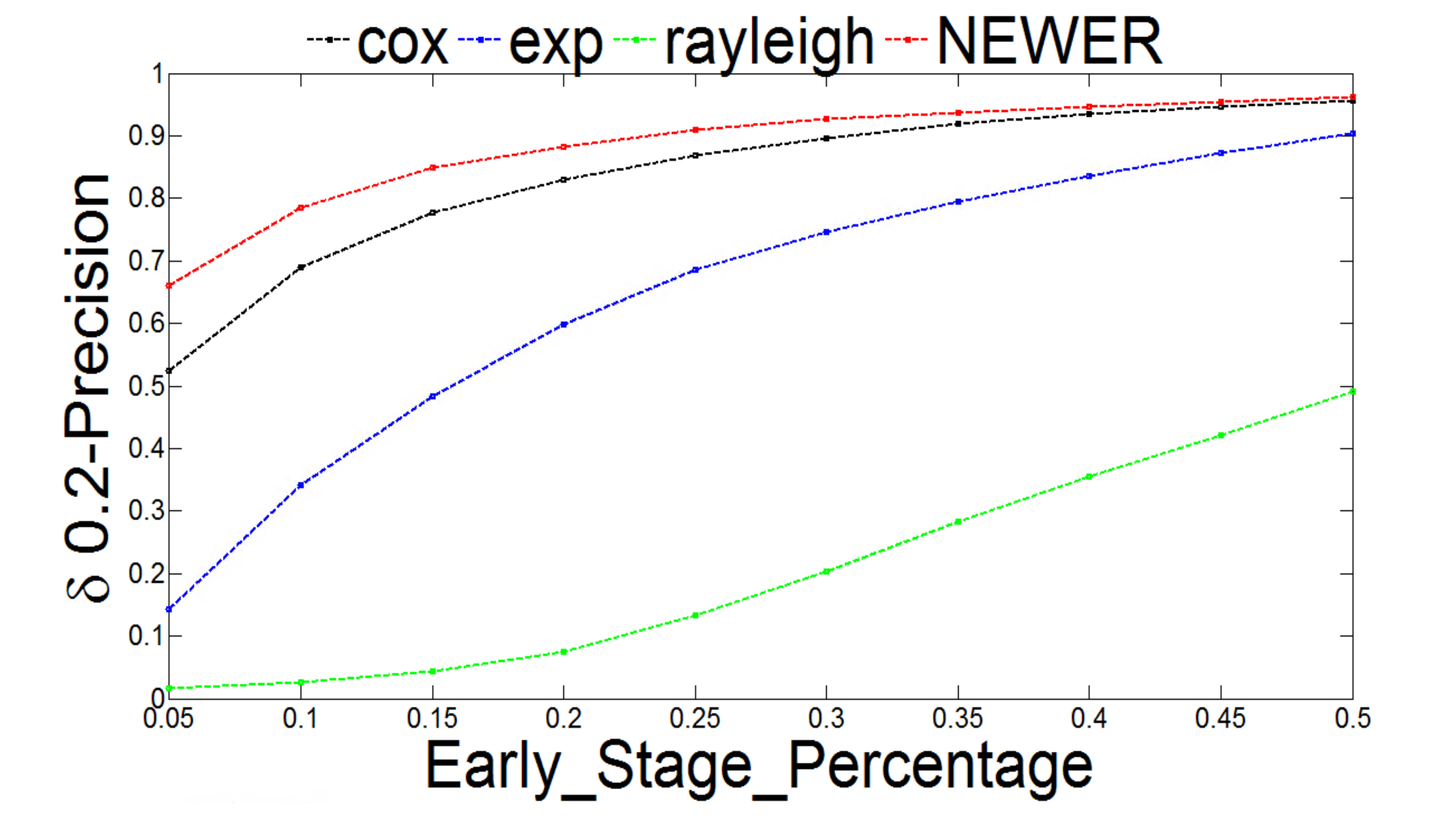}
    \vspace{-0.325in}

    \caption{Cascading process prediction accuracy of different methods under different early stage percentage settings.}
    \label{fig:exp_early_stage}
    \vspace{-0.205in}
\end{figure}

The ultimate purpose of this paper is to predict the cascading process. For each cascade, we use $\delta t$ to represent the early stage window and $\hat{t}$ to represent its ending time. Then we use the cascade information during $[0,\delta t]$ to predict the cascading process during $[\delta t, \hat{t}]$. At any time $t\in[\delta t, \hat{t}]$, we calculate whether the predicted cascade size at $t$ is within the $\sigma$ tolerance of the groundtruth size at $t$. Then we calculate the $\delta\sigma$-Precision by integrating $t$ to describe the prediction accuracy for this cascading process. Finally, we average the $\delta\sigma$-Precision for all cascades and show the results in Figure \ref{fig:exp_early_stage}. Here, we vary the early stage percentage (i.e. $\delta t/\hat{t}$) from 0 to $50\%$, and discover that in all the settings of early stage percentage, NEWER always carries out the best performances in cascading process prediction. More over, the advantage of NEWER is more clear in smaller early stage percentage. When we set the early stage to be $15\%$ of the whole cascade duration, we can get the $\delta 0.2$-Precision of $0.849$. That means that we can correctly predict the cascade sizes at $84.9\%$ time points, which indicates that the cascading process is predictable and the proposed method is adequate and superior in cascading process prediction. Furthermore, changing the precision tolerance value $\sigma$ will not affect the relative results of all the methods in our experiments, and the precision value will be smaller when setting $\sigma$ smaller. For abbreviation, we only report the results of $\sigma=0.2$, which is a reasonable tolerance in most application scenarios.

\subsection{Out-of-sample Prediction}

In real applications, the interaction information between nodes is not always available, which makes some nodes' behavioral dynamics cannot be directly derived by maximum likelihood estimation from data. We call these nodes as \emph{out-of-sample nodes}. This is the main reason why we propose NEWER to incorporate the covariates of behavioral dynamics. In order to evaluate the performance of NEWER in handling this case, we simulate the scenario by hiding the interaction information of randomly selected $10\%$ users as out-of-sample users, and then predict the final sizes of the cascades that these users involved in early stages.

\begin{figure}[htbg]
    \vspace{-0.135in}
    \centering
    \includegraphics[width=0.5\textwidth]{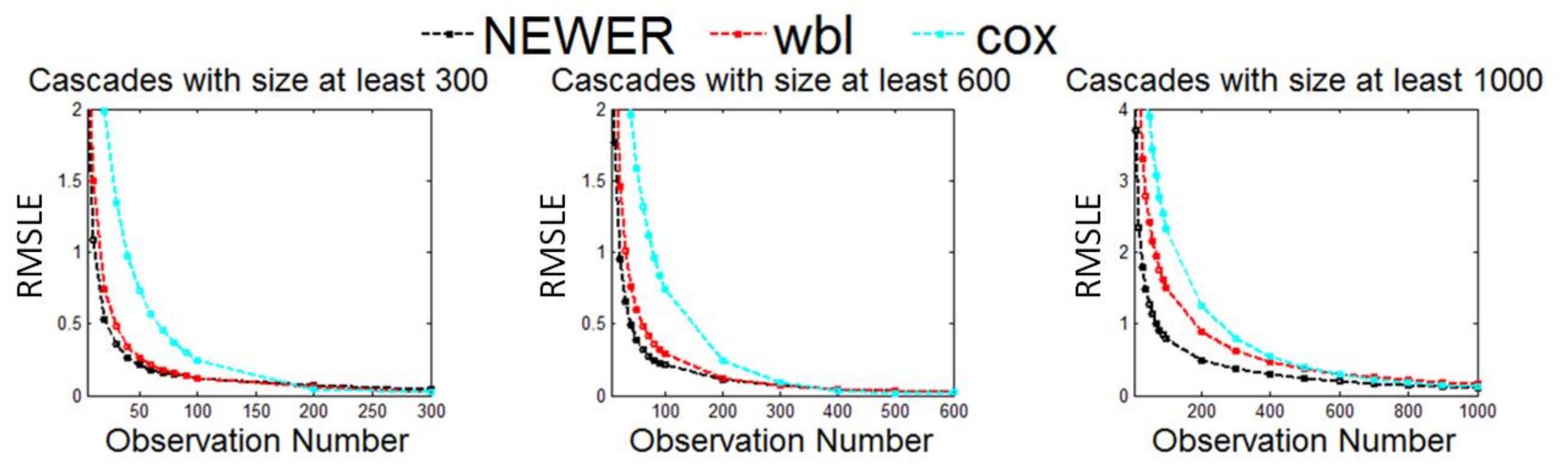}
    \vspace{-0.425in}

    \hspace{0.0002in}
    \caption{Prediction result by unknown users}
    \label{fig:unknown_user_prediction}
    \vspace{-0.135in}
\end{figure}

In Cox model, the scale parameters in behavioral dynamics of out-of-sample users can be regressed by the covariates. For the shape parameter, we calculate the average value of shape parameters in observed users and apply this value to the shape parameters of out-of-sample users. In NEWER model, both of shape and scale parameters can be regressed by covariates with the learned $\beta$ and $\gamma$. We also employ the standard Weibull Regression (Wbl) as a basline, which can be derived by simply setting $\mu$ and $\eta$ to be 0 in Equation \ref{eq:objectivefunction}. Then we use the averaged shape and scale parameters of observed users as the parameters of out-of-sample users.

\begin{figure}[htbg]
    \vspace{-0.135in}
    \centering
    \includegraphics[width=0.5\textwidth]{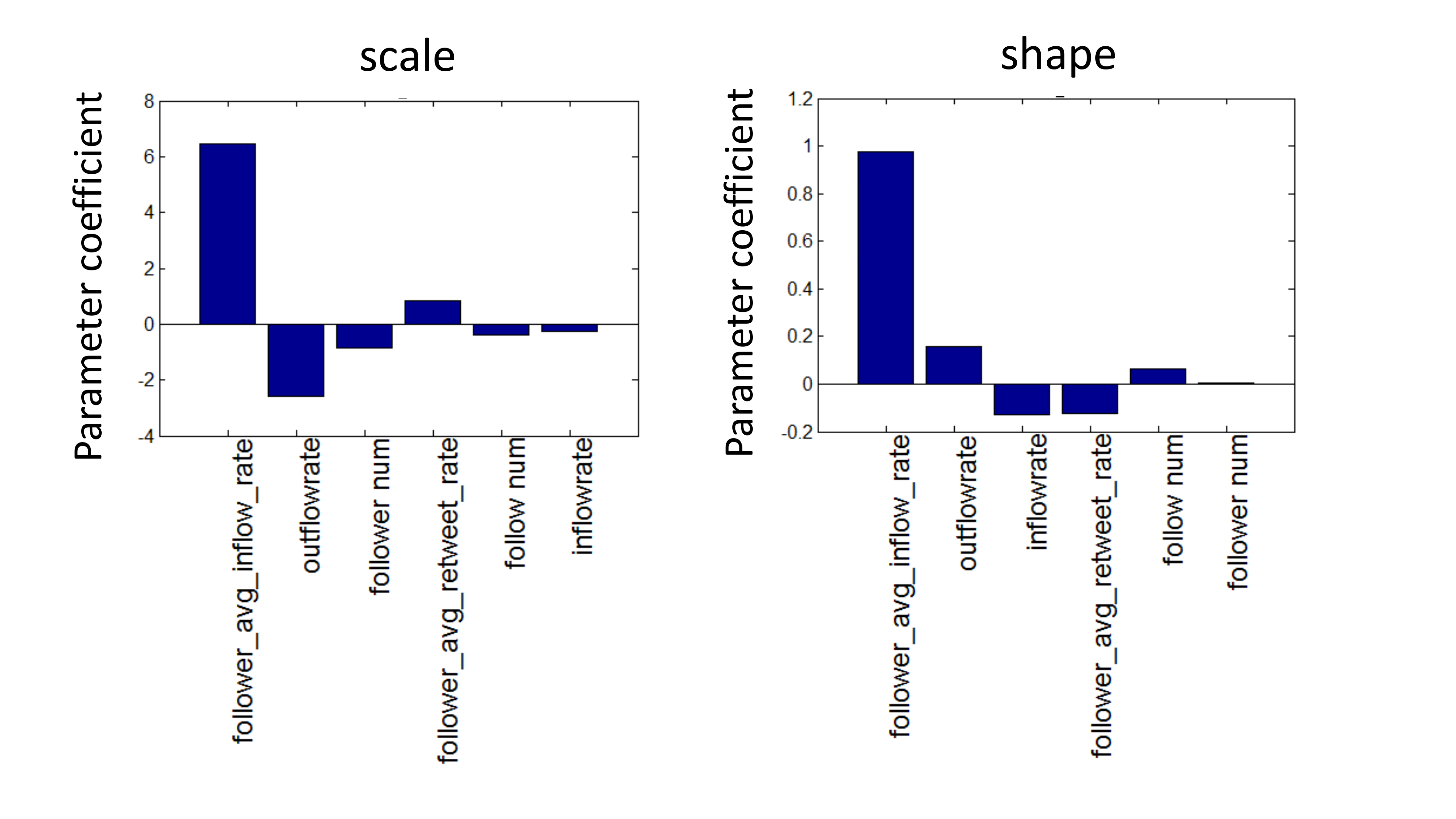}
    \vspace{-0.425in}

    \caption{Parameter coefficients.}
    \label{fig:exp_par_coe}
    \vspace{-0.21in}
\end{figure}

As shown in Figure \ref{fig:unknown_user_prediction}, the NEWER model can significantly and consistently outperform Cox and Wbl models in out-of-sample prediction, which demonstrates that the discovered covariates from behavioral features of a user's networked neighbors can effectively predict the user's behavioral dynamics. Also, we visualize the regression coefficients $\beta$ and $\gamma$ in Figure \ref{fig:exp_par_coe}. It can be observed that the behavioral features of a user's followers plays more important roles in predicting both scale and shape parameters for the user, while the user's structural features are less important.

\input

\section{Conclusions}

In this paper, we raise an important and interesting question: beyond predicting the final size of a cascade, can we predict the whole cascading process if the early stage information of cascades is given? In order to address this problem, we propose to uncover and predict the macro cascading process with micro behavioral dynamics. Through data-driven analysis, we find out the common principles and important patterns laying in behavioral dynamics, and propose a novel NEWER model for behavioral dynamics modeling with good interpretability and generality. After that, we propose a scalable method to aggregate micro behavioral dynamics into macro cascading processes. Extensive experiments on a large scale real data set demonstrate that the proposed method achieves the best results in various cascading prediction tasks, including cascade size prediction, outbreak time prediction and cascading process prediction.

\vspace{-0.05in}

\section*{Appendix: Proof of Theorem 1}
\vspace{-0.5em}
\noindent\begin{proof}
\small
    	It's evident that both $G_2(\beta, \lambda)$ and $G_3(\gamma, k)$ has global minimum value. Next we prove that $G_1(\lambda, k)$ also has global minimum value, or to prove $\log{L(\lambda, k)}$ has global maximum value.

		Let $\lambda_i'=\lambda_i^{-k_i}$, $\log{L'(\lambda', k)}=\log{L(\lambda, k)}=\sum_{i=1}^N l_i'(\lambda_i',k_i)$ where $l_i'(\lambda_i',k_i)=m_i\log{k_i}+(k_i-1)\sum_{j=1}^{m_i}{\log{T_{i,j}}}+m_i\log{\lambda_i'}-\lambda_i'\sum_{j=1}^{m_i}{T_{i,j}^{k_i}}$, the partial derivatives of the $l_i'$ are given by:
    \vspace{-0.105in}
		\begin{align}
			&\frac{\partial{l_i'}}{\partial{\lambda_i'}}=\frac{m_i}{\lambda_i'}-\sum_{j=1}^{m_i}{T_{i,j}^{k_i}},~~\frac{\partial^2{l_i'}}{\partial{\lambda_i'^2}}=-\frac{m_i}{\lambda_i'^2}<0\\
			&\frac{\partial{l_i'}}{\partial{k_i}}=\frac{m_i}{k_i}+\sum_{j=1}^{m_i}{\log{T_{i,j}}}-\lambda_i'\sum_{j=1}^{m_i}{T_{i,j}^{k_i}\log{T_{i,j}}}\\
			&\frac{\partial^2{l_i'}}{\partial{k_i^2}}=-\frac{m_i}{k_i^2}-\lambda_i'\sum_{j=1}^{m_i}{T_{i,j}^{k_i}\left(\log{T_{i,j}}\right)^2}<0
		\end{align}
		Since $\frac{\partial^2{l_i'}}{\partial{\lambda_i'^2}}<0$ and $\frac{\partial^2{l_i'}}{\partial{k_i^2}}<0$, the conditional marginal posterior densities of parameters $\lambda_i'$ and $k_i$ are log-concave. Moreover, when $0<k_i<1, 0<\lambda_i'<\min\left(\frac{m_i}{\sum_{i=1}^{m_i}T_{i,j}^{k_i}},\frac{1}{\sum_{j=1}^{m_i}{T_{i,j}\log{T_i}}}\right)$,
    \vspace{-0.105in}
		\begin{align}
			&\frac{\partial{l_i'}}{\partial{\lambda_i'}}=\frac{m_i}{\lambda_i'}-\sum_{j=1}^{m_i}{T_{i,j}^{k_i}}\geq \frac{m_i}{\frac{m_i}{\sum_{i=1}^{m_i}T_{i,j}^{k_i}}}-\sum_{j=1}{m_i}{T_{i,j}^{k_i}}=0\\
			&\frac{\partial{l_i'}}{\partial{k_i}}=\frac{m_i}{k_i}+\sum_{j=1}^{m_i}{\log{T_{i,j}}}-\lambda_i'\sum_{j=1}^{m_i}{T_{i,j}^{k_i}\log{T_{i,j}}} \nonumber\\
			&\geq \frac{m_i}{k_i}+\sum_{j=1}^{m_i}{\log{T_{i,j}}}-\frac{\sum_{j=1}^{m_i}{T_{i,j}^{k_i}\log{T_{i,j}}}}{\sum_{j=1}^{m_i}{T_{i,j}\log{T_i}}}\geq m_i + \sum_{j=1}^{m_i}{\log{T_{i,j}}} - 1 > 0\nonumber
		\end{align}
		when $k_i \geq \max\left(1, \frac{m_i}{\lambda_i'\sum_{j=1}^{m_i}{T_{i,j}\log{T_{i,j}}} - \sum_{j=1}^{m_i}{\log{T_{i,j}}}} \right)$ and $\lambda_i'\geq \max\left(1, \frac{m_i}{\sum_{j=1}^{m_i}T_{i,j}^{k_i}}\right)$,
    \vspace{-0.105in}
		\begin{align}
			&\frac{\partial{l_i'}}{\partial{\lambda_i'}}=\frac{m_i}{\lambda_i'}-\sum_{j=1}^{m_i}{T_{i,j}^{k_i}}\leq \frac{m_i}{\frac{m_i}{\sum_{i=1}^{m_i}T_{i,j}^{k_i}}}-\sum_{j=1}{m_i}{T_{i,j}^{k_i}}=0\\
			&\frac{\partial{l_i'}}{\partial{k_i}}=\frac{m_i}{k_i}+\sum_{j=1}^{m_i}{\log{T_{i,j}}}-\lambda_i'\sum_{j=1}^{m_i}{T_{i,j}^{k_i}\log{T_{i,j}}}\nonumber\\
			&\leq \frac{m_i}{\frac{m_i}{\lambda_i'\sum_{j=1}^{m_i}{T_{i,j}\log{T_{i,j}}} - \sum_{j=1}^{m_i}{\log{T_{i,j}}}}} +\sum_{j=1}^{m_i}{\log{T_{i,j}}}-\lambda_i'\sum_{j=1}^{m_i}{T_{i,j}^{k_i}\log{T_{i,j}}}\nonumber\\
			&= \lambda_i'\sum_{j=1}^{m_i}{T_{i,j}\log{T_{i,j}}} - \sum_{j=1}^{m_i}{\log{T_{i,j}}}+\sum_{j=1}^{m_i}{\log{T_{i,j}}}-\lambda_i'\sum_{j=1}^{m_i}{T_{i,j}^{k_i}\log{T_{i,j}}}\nonumber\\
			&<0
		\end{align}
 		which means there should be a global maximum of $l_i'$, so does $\log L$.
	\end{proof}

%

\bibliographystyle{abbrv}
\bibliography{sigproc}  
%
%
\balancecolumns

\end{document}